\documentclass[10pt]{extarticle}
\usepackage{spconf,amsmath,graphicx,hyperref,enumitem,xcolor,soul,amsmath, amssymb,graphicx,siunitx}

\usepackage{etoolbox}
\apptocmd{\thebibliography}{\ninept}{}{}

\usepackage{booktabs}  
\usepackage[para]{footmisc}
\usepackage{graphicx}

\title{RESTORE: REal-time Steerable Music resTORation and bandwidth Extension via stem disentanglement}

\name{Meiying Chen, Benjamin R. Thompson and Michael C. Heilemann}
\address{University of Rochester\\
	Department of Electrical and Computer Engineering, Rochester NY, USA}

\begin{document}

\maketitle

\begin{abstract}
\end{abstract}
Neural methods for audio restoration are typically framed as rigid mappings from degraded inputs to single clean outputs, enforcing decisions about what audio content is removed, and potentially adding unwanted content to the restored signal. Because what constitutes a restored audio signal is subjective, we introduce RESTORE, a framework that formulates audio restoration as a six-source semantic decomposition to allow for real-time interactive user control over the process. By expanding a pretrained HTDemucs backbone, a single forward pass disentangles a degraded mixture into vocals, music, broadband hiss, impulsive transients, and an unmodeled residual, while jointly synthesizing a high-frequency extension. Users may steer the restoration by adjusting stem gains, ensuring generative content remains isolated and auditable. 
RESTORE improves audio quality on diverse historical recordings compared to baselines,
lowering Fr\'echet Audio Distance (FAD) (12.13 VGGish; 0.92 CLAP) and delivering aesthetic steerability (Spearman $\rho \ge 0.91$) at $50\times$ real-time on a single GPU\footnote{Code and audio samples are available at \url{https://melissachen15.github.io/restore-audio-demo/}}.




\section{Introduction}
\label{sec:intro}

The restoration of historic audio such as shellac discs, presents a unique challenge at the intersection of signal processing and preservation \cite{valimaki2008digital}. These recordings suffer from heavy band-limiting, broadband hiss, impulsive transients and other degradations imparted by the recording equipment, the manufacturing process, the playback medium and age-related wear \cite{valimaki2008digital, juanpere2022realistic}. 
Neural enhancement models provide a powerful solution, but they typically remove all of the noise they are able to identify rather than allowing the practitioner to adjust the level of intervention.
\cite{li2020learning, kandpal2022music}. 
 
The definition of what constitutes a restored audio signal is highly subjective, and the complete removal of all noise may sound inauthentic and unnatural compared to less complete removal \cite{valimaki2008digital}. To address this gap, recent research explores generative models for enhancement and blind audio bandwidth extension. While single-pass systems exist for speech 
\cite{liu2022voicefixer},
frameworks targeting historical music, such as U-Nets \cite{moliner2022two}, GANs \cite{moliner2022behm} and diffusion models 
\cite{moliner2024diffusion, moliner2024buddy} rely on cascaded pipelines that sequentially perform separation, denoising, and extension. 
These cascaded designs may lead to compounding errors including hallucinated vocal content, as early separation mistakes propagate through to the final generative stage. Furthermore, the high computational cost of diffusion prevents real-time use \cite{vsvento2026music, moliner2024diffusion}, and models trained on 
solo singing and piano recordings struggle to generalize to complex multi-voice mixtures \cite{moliner2024diffusion}. Conversely, music source separation models \cite{rouard2023hybrid, luo2023music} prioritize real-time efficiency, and can isolate complex musical mixtures with very low latency. However, these models expect largely noise-free input signals and only separate existing content. They lack the generative capacity needed to synthesize the missing high-frequency bandwidth lost in historical media. 

To combine high-quality processing with real-time control, we introduce RESTORE: REal-time, Steerable music resTORation and bandwidth Extension. 
We adapt Hybrid Transformer Demucs (HTDemucs) \cite{rouard2023hybrid}, a dual-domain
music source separation network, by replacing the final decoder layer of each branch with six source-specific output heads, all produced in a single forward pass: two clean content stems (vocals, music), three degradation stems (hiss, transients, residual), and a high-frequency extension (HFE) stem synthesized under a targeted adversarial loss. A mixture-consistency penalty enforces an additive decomposition, ensuring degradation is accounted for in the noise stems rather than discarded. The contributions of this paper are:

\noindent\textbf{Semantic Disentanglement for Controllability}: We repurpose a separation network into a semantic noise decomposer with aesthetic control.

\noindent\textbf{Adversarial Bandwidth Extension}: Isolating the generated bandwidth into an independent stem allows for targeted adversarial training without corrupting the original signal.

\begin{figure*}[t]
  \centering
\includegraphics[width=.95\textwidth]{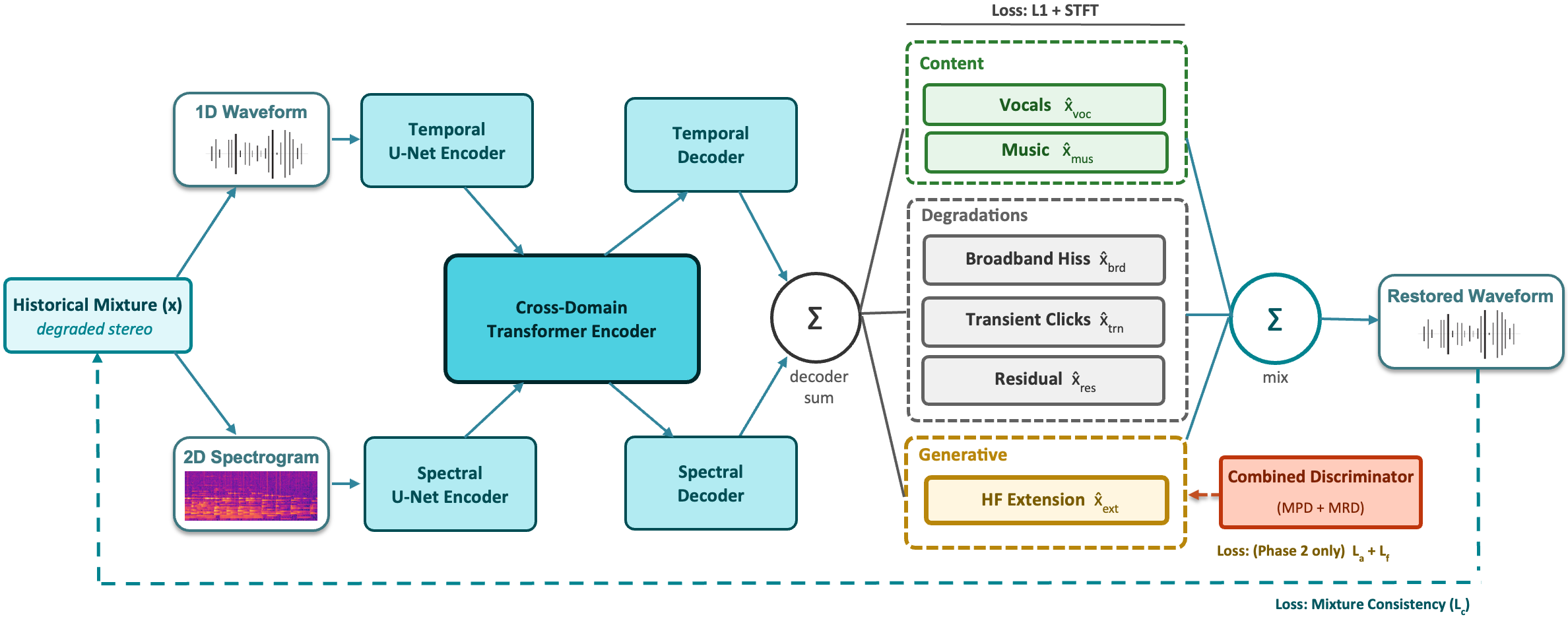}
\caption{System overview: a degraded historical mixture is processed by a dual-domain generator with a cross-domain transformer bottleneck, then decoded and split into six semantically targeted stems. All stems are supervised via L1 and STFT losses with a mixture consistency penalty enforcing the sum back to the input. A discriminator is applied only to the high-frequency extension stem. The restored waveform is reconstructed by summing desired stems with independent, user-defined gains.}
\label{fig:system_overview}
\end{figure*}

\noindent\textbf{Universal Generalization}: Unlike previous models that overfit to solo sources, RESTORE natively generalizes to complex \SI{44.1}{\kilo \hertz} stereo polyphonic mixtures.


\section{Method}
\label{sec:method}

A complete system overview is shown in Fig.~\ref{fig:system_overview}. We build RESTORE based on HTDemucs architecture \cite{rouard2023hybrid}, which employs a bi-U-Net structure that operates in both the temporal and spectral domains for music source separation. Its outermost layers consist of standard convolutions, and the bottleneck is a cross-domain Transformer encoder. This encoder interleaves self-attention within each domain and cross-attention between domains, allowing it to efficiently capture complex musical structures.

In its standard configuration, HTDemucs separates clean music into four fixed stems: drums, bass, vocals, and other. Since historic audio is degraded and may lack this predictable structure, we expand the network output to six semantically meaningful, specialized stems: Vocals, Music, Broadband Noise, Transient Noise, Residual, and HFE. Isolating HFE into a dedicated stem offers users the flexibility to scale or discard the synthesized frequencies. Because synthesizing missing harmonics is fundamentally a generative task and isolating existing signals is discriminative, decoupling these objectives into distinct heads helps prevent conflicting learning signals during training. A standard HTDemucs model may group clicks and broadband hiss into the drums stem, due to their respective similarities to impulsive strikes and signals with noise-like features such as snare drums. We leverage this behavior by mapping our noise stems to the pretrained drums weights. To prevent these copied heads from emitting large, incorrectly scaled signals at initialization, we attenuate the transient and extension heads to keep their outputs small while the network adapts as shown in Table \ref{tab:init_scale}.

\begin{table*}[t]
\centering
\vspace{-6.89pt}
\caption{Output Head Initialization, Scaling, and Ground-Truth Targets}
\label{tab:init_scale}
\begin{tabular}{cccc}
\hline
\textbf{New Stem} & \textbf{Pretrained Donor Source} & \textbf{Init. Scale} & \textbf{Ground-Truth Signal} \\
\hline
Vocals & Vocals & 1.0 & Lowpass-filtered clean vocals \\
Music & Other & 1.0 & Lowpass-filtered clean accompaniment/music \\
Broadband & Drums & 1.0 & Hiss signal injected during simulation \\
Transient & Drums & 0.075 & Click and thump signals injected during simulation \\
Extension & Other & 0.05 & (Clean mixture) $-$ (Lowpass mixture) $=$ (Missing high frequencies) \\
Residual & Bass & 1.0 & Other \\
\hline
\end{tabular}
\end{table*}

To construct paired training data, we simulate the degradations of historical recordings on clean, full-band audio \cite{juanpere2022realistic,valimaki2008digital,moliner2022two}. All degradations are applied online, so no two batches see the same corruption of a given clip. 
Following \cite{moliner2024diffusion}, we apply a piece-wise linear frequency-domain filter, by first defining a breakpoint with a frequency $f_0\in [0.5, 1.5]$ \SI{}{\kilo \hertz} at \SI{0}{\deci\bel}. Additional breakpoints are defined at frequencies of $f_0/4$, $f_0/2$, $2f_0$ and $4f_0$, with magnitudes determined by connecting line segments with slopes $\in[-12,-2]$ \SI{}{\deci\bel\per oct.}. To impose high- and low-pass filtering, gains are rolled off from the first and last breakpoints at \SI{80}{\decibel\per oct.}. Additionally, a Butteworth low-pass filter was applied using the $\operatorname{buttord\cite{scipy_buttord}}$ function in SciPy, with passband and stopband frequencies of \SI{2.8}{\kilo\hertz} and \SI{4.4}{\kilo\hertz} respectively (randomized within -20 to +15$\%$) to mimic the response of historic equipment.


Next, noise from the Gramophone Record Noise Dataset \cite{moliner2022two} is injected and scaled to SNRs sampled from an empirical distribution measured on 30 digitized 78 RPM discs (15.2--\SI{52.3}{\deci\bel}) \cite{great78}, with up to \SI{4}{\deci\bel} jitter and a \SI{5}{\deci\bel} floor to prevent the model from overfitting to discrete SNR levels. Clicks follow a log-uniform rate of 0.5--5 clicks/s, whose median (1.58) matches measurements from real \SI{78}{RPM} discs; 
bursts below \SI{60}{hertz} are repeated at intervals of \SI{0.769}{\second}, modeling an eccentric or warped pressing where the stylus meets the same defect each rotation.
Distortion \cite{valimaki2008digital} and groove echo are omitted, as they did not impact preliminary experiments.

Synthesizing the HFE stem requires the network to fabricate missing content, and L1 and STFT losses alone tend to yield robotic and over-smoothed high frequencies \cite{kumar2019melgan}. Therefore, RESTORE employs a two-stage adversarial training schedule. In the first stage, we freeze the transformer bottleneck and train without discriminator to establish basic semantic separation. In the second stage, we unfreeze the transformer and activate a Multi-Period \cite{kong2020hifi} and a Multi-Resolution Discriminator \cite{lee2022bigvgan}. We penalize the generator via LSGAN ($L_{a}$) \cite{mao2017least} and feature-matching ($L_{f}$) \cite{kumar2019melgan} losses. We apply this discriminator exclusively to the extension stem to avoid corrupting the stems with surviving original content.

\subsection{Training Objectives and Mixture Consistency}
Each simulated degradation maps to a ground-truth stem as in Table \ref{tab:init_scale}. 
We extend the HTDemucs L1 waveform loss into:

\vspace{-12pt}
\begin{equation}
L = \sum_{s} w_s \left[ \text{L1}_s + \Psi_s \right] + \lambda_{c} L_{c} + \lambda_{r} L_{r}
+ \lambda_{a} L_{a} + \lambda_{f} L_{f}
\end{equation}
\vspace{-10pt}

\noindent where $w_s = 1.0$ for all stems but HFE, which is held at 0.1 until the discriminator engages, then ramped to 1.0; $\lambda_{c} = 0.1$; $\lambda_{r} = 10^{-3}$, $\lambda_{a} = 1.0$; $\lambda_{f} = 2.0$. $L_{a}$ and $L_{f}$ apply to the HFE stem and are enabled in phase 2 only. $\text{L1}_s$ is computed on pre-emphasised waveforms; $\Psi_s$ is a multi-resolution STFT loss over FFT sizes 512, 1024 and 2048. Log-magnitude spectral losses are mathematically unstable on silent targets, and our transient and residual stems are predominantly silence. Accordingly, we split the bins by target energy: active bins receive the log-magnitude penalty, while near-zero bins receive a linear L1 penalty at weight 0.1. This arrangement penalizes hallucinated noise in silent regions. We also require the five band-limited stems to add back up to the mixture, so every part of the input is accounted for and errors cannot be quietly dumped into the unconstrained HFE stem. The L1 penalty $L_c$ minimizes the difference between the input mixture and the sum of all 5 predicted stems excluding HFE. The residual stem acts as a catch-all for unmodeled content. A light penalty $L_{r}$ on its energy forces it to default to silence. In phase two, adversarial penalties $L_{a}$ and $L_{f}$ are activated exclusively on the HFE stem to drive the synthesis of realistic high frequencies.




\section{Experimental Setup}
\label{sec:exps}

The training pipeline draws from a pool of 3,700 clean vocal stems and 1,790 instrumental stems at 44.1\,kHz, 16-bit stereo. Vocal stems are sampled from MUSDB18-HQ \cite{MUSDB18HQ} (50\%) and VocalSet \cite{wilkins2018vocalset} (50\%); music stems from MUSDB18-HQ (40\%), MAESTRO \cite{hawthorne2018enabling} (25\%), MusicNet \cite{thickstun2016learning} (20\%), and URMP \cite{li2018creating} (15\%). Ten MUSDB18-HQ songs are held out for validation. During training, we extract random 6-second segments and mix them with gains between \SI{-6}{\deci\bel} and \SI{+3}{\deci\bel}. 
10\% of mixtures are vocal-only and 10\% are instrumental-only. Degradation is applied on-the-fly per sample.

For real-world evaluation, we curate a specialized test set of 78 RPM recordings from \cite{great78}. For direct comparison with our primary baseline BABE-2 \cite{moliner2024diffusion}, we randomly sample 30 in-domain recordings from its evaluation set: Enrico Caruso (10), Nellie Melba (10), and solo piano (10). The Caruso and Melba pieces are vocal-dominant with very light accompaniment, and the piano pieces are purely instrumental. To probe robustness beyond that domain, we further randomly sampled 18 additional diverse historical recordings. This generalization subset has 10 categories including flamenco, jazz, orchestral, military band, solo violin/guitar, blues and others. All historic recording are trimmed to 60 seconds.

RESTORE was trained on a NVIDIA A100 GPU using the two-stage process described above. Phase one spans 75k steps with a batch size of 6. Phase two contains 200k steps, with a batch size reduced to 2 to accommodate the increased GPU memory demands of the adversarial network. We optimize the model using AdamW on a cosine learning rate schedule, which includes a 3k step warmup to a peak learning rate of $10^{-4}$ during phase one, and continuing through phase two. The training process completed in 56 hours. 

\subsection{Baselines and Evaluation Metrics}
RESTORE is benchmarked against industry-standard software and a state-of-the-art deep learning model. iZotope RX 12 Advanced (v12.0.0.1410) \cite{izotope2026rx} is used as a commercial baseline with the sequence recommended for archival material (De-click, De-crackle, De-hum, Spectral De-noise, and Spectral Recovery) representing the conventional multi-pass DSP workflow adopted by many audio engineers. The deep learning baseline, BABE-2 \cite{moliner2024diffusion} restores bandwidth and coloration but does not remove noise, so its authors deploy it behind a preprocessing cascade that we reproduce: a two-stage U-Net denoiser \cite{moliner2022two} suppresses hiss and clicks, and for vocal recordings, HT-Demucs \cite{rouard2023hybrid} isolates the voice before the diffusion pass. Piano and instrumental recordings use the denoiser and BABE-2 without the separation step. We use the authors' released checkpoints: the singing model for vocals, and the MAESTRO model for instrumentals. We omit the per-singer fine-tuning, as the generic prior better reflects how the model would be used in practice than a timbre-matched one. Finally, HT-Demucs is also our own backbone, so the comparison isolates the effect of our training.

On the paired synthetic test set we compute the Scale-Invariant Signal-to-Distortion Ratio (SI-SDR) \cite{le2019sdr} to measure core separation and denoising accuracy. We also report ViSQOL (Virtual Speech Quality Objective Listener, audio mode) \cite{chinen2020visqol} for the overall perceptual quality. 
Because historical recordings lack clean ground truth, we use reference-free Fr\'echet Audio Distance (FAD) \cite{kilgour2018fr}. To capture both acoustic characteristics and semantic content, we compute FAD using VGGish \cite{hershey2017cnn} and CLAP \cite{wu2023large} feature extractors. 
We define \emph{steerability} as the Spearman correlation between a stem's applied gain (swept from 0 to 1) and that stem's realized RMS energy in the mixture. A coefficient near 1 indicates the head is controllable and not entangled with the others. 
For the voice stem, we measure \emph{noise leak}: the residual noise energy in the vocal estimate. 
We also report \emph{music bleed}, the fraction of a noise stem's energy explained by projection onto the clean music. A value near zero means muting that stem removes almost no music. 
Finally, we report the median real-time factor (RTF), defined as processing time divided by audio duration, on a single NVIDIA A100 GPU.

\begin{table}[t]
\centering
\caption{Results on the synthetic test set with paired clean references ($n{=}30$). \emph{Voice stem}: scored against the band-limited clean vocal (LP: low-passed to match bandwidth). \emph{Restored recording}: full-band output vs. clean mixture. FAD-V/C: VGGish/CLAP embeddings.}
\label{tab:synthetic}
\setlength{\tabcolsep}{3pt}
\resizebox{\columnwidth}{!}{%
\begin{tabular}{lccccc}
\toprule
Method & SI-SDR$\uparrow$ & ViSQOL$\uparrow$ & FAD-V$\downarrow$ & FAD-C$\downarrow$ & Leak (dB)$\downarrow$ \\
\midrule
\multicolumn{6}{l}{\textit{Restored recording (full-band)}} \\
Unprocessed & $-1.28$ & 1.62 & 9.07 & 0.86 & -- \\
HT-Demucs & $-1.32$ & 1.65 & 8.74 & 0.83 & -- \\
U-Net\footnotemark{}\ & $-1.09$ & 1.84 & 6.81 & 0.81 & -- \\
iZotope & $-3.91$ & 1.83 & 9.19 & 0.93 & -- \\
Ours & \textbf{0.76} & \textbf{2.86} & \textbf{4.11} & \textbf{0.40} & -- \\
\midrule
\multicolumn{6}{l}{\textit{Voice stem (band-limited)}} \\
BABE-2 (LP) & 1.11 & 4.22 & 5.99 & 0.82 & $-2.86$ \\
HT-Demucs (LP) & 4.21 & 4.39 & 4.00 & 0.60 & $-5.38$ \\
Ours & \textbf{7.65} & \textbf{4.52} & \textbf{3.43} & \textbf{0.53} & $\mathbf{-8.36}$ \\
\midrule
\multicolumn{1}{l}{\textit{Control (ours)}}
 & \multicolumn{2}{c}{Steerability$\uparrow$} & \multicolumn{3}{c}{Music bleed$\downarrow$} \\
HFE & \multicolumn{2}{c}{0.973} & \multicolumn{3}{c}{--} \\
Transient & \multicolumn{2}{c}{0.910} & \multicolumn{3}{c}{0.0003} \\
Broadband & \multicolumn{2}{c}{0.933} & \multicolumn{3}{c}{0.0017} \\
\bottomrule
\end{tabular}}
\end{table}
 
\begin{table}[t]
\centering
\footnotesize
\caption{Results on Historic recordings (no clean reference). Songs: blues, Caruso, country, flamenco, Melba, tango, yodel; instrumental: jazz, military, orchestra, piano, violin/guitar solo.}
\label{tab:historical}
\setlength{\tabcolsep}{2.7pt}
\begin{tabular}{@{}lcccccc@{}}
\toprule
 & \multicolumn{2}{c}{Songs} & \multicolumn{2}{c}{Instrumental} & & \\
\cmidrule(lr){2-3}\cmidrule(lr){4-5}
Method & FAD-V & FAD-C & FAD-V & FAD-C & RTF & \#Params(M)\footnotemark \\
\midrule
\multicolumn{7}{l}{\textit{Restored recording (full band)}} \\
Unprocessed & 22.80 & 1.08 & 19.72 & 1.18 & -- & -- \\
HT-Demucs & 19.10 & 0.99 & 15.33 & 1.13 & 0.0200 & 41.98 \\
U-Net\footnotemark[\numexpr\value{footnote}-1\relax] & 14.09 & 0.97 & 12.69 & 1.29 & 0.9864 & 71.41 \\
iZotope & 13.10 & 1.08 & 14.23 & 1.27 & $\sim$2--3\footnotemark & -- \\
Ours (HFE off) & 12.71 & 0.94 & 9.06 & \textbf{1.09} & \textbf{0.0195} & 41.99 \\
Ours (HFE on) & \textbf{12.13} & \textbf{0.92} & \textbf{8.60} & 1.15 & \textbf{0.0195} & 41.99 \\
\midrule
\multicolumn{7}{l}{\textit{Voice stem (band-limited)}} \\
BABE-2 (LP) & 11.74 & \textbf{1.12} & -- & -- & 26.7896 & 154.8$^\S$ \\
Ours & \textbf{10.59} & 1.13 & -- & -- & \textbf{0.0195} & 41.99 \\
\bottomrule
\end{tabular}
\end{table}

{\renewcommand\thefootnote{}%
\footnotetext{%
  \textsuperscript{\the\numexpr\value{footnote}-2\relax}\,BABE-2 stage-1 denoiser.\quad
  \textsuperscript{\the\numexpr\value{footnote}-1\relax}\,Size includes the U-Net denoiser, HT-Demucs, and a 41.4M-parameter diffusion model.\quad
  \textsuperscript{\the\value{footnote}}\,Rough estimate, iZotope RX GUI on a MacBook Air M2.}}
\section{Results and Conclusion}
\label{sec:results}


For full-band restoration on synthetic data, our model outperforms all baselines on every metric, achieving the highest SI-SDR (0.76\,dB) and ViSQOL (2.86) and half the FAD-CLAP of the strongest baseline (0.40 vs.\ 0.81) as shown in Table~\ref{tab:synthetic}. HT-Demucs shows only a small improvement (FAD-CLAP 0.83 vs.\ 0.86), despite sharing our architecture and size. This indicates that our performance gains result from our task formulation and training strategy rather than model capacity. At a matched bandwidth, our isolated voice stem achieves the best SI-SDR (\SI{7.65}{\deci\bel} vs. HTDemucs's \SI{4.21}{\deci\bel}), and leaves the lowest noise in voiced frames (\SI{-8.36}{\deci\bel} vs. \SI{-5.38}{\deci\bel} and \SI{-2.86}{\deci\bel}). On historic recordings, our model achieves the lowest FAD across both song and instrumental categories for full-band restoration. It outperforms baselines even with the generative HFE disabled, confirming that separation drives the gains and leaving HFE as a controllable feature for the archivist to enable. On the isolated voice stem, RESTORE remains competitive with the cascaded BABE-2, surpassing it on FAD-V with comparable FAD-C.


A single forward pass produces all six stems at an RTF of 0.0195, about 50$\times$ faster than the U-Net and over 1,300$\times$ faster than BABE-2, with 3.7$\times$ fewer parameters.
Furthermore, output stems respond monotonically to their gain (Spearman 0.91--0.97), so each artifact type can be attenuated or restored independently.
Music bleed into the transient and broadband stems is near zero (0.03\% and 0.17\% of their energy), so muting either stem removes almost no music.

RESTORE demonstrates that a multi-task separation network can simultaneously serve as a semantic decomposer and a generative bandwidth-extension engine. This unified, single-pass approach grants users real-time, post-hoc control over each artifact class, while its exact-additivity target factorization mitigates the hallucination risks of cascaded pipelines. 
Future work includes developing independent HFE heads for vocal and music stems, and conducting listening studies to validate perceptual quality, archival fidelity, and steerability on real recordings.


\section{Acknowledgements}
This work used Delta at NCSA through allocation CIS261364 from the Advanced Cyberinfrastructure Coordination Ecosystem: Services \& Support (ACCESS) program, which is supported by NSF grants \#2138259, \#2138286, \#2138307, \#2137603, and \#2138296. The work was supported by a research initiation grant from the SoundSpace Institute.


\bibliographystyle{IEEEbib}
\bibliography{refs}
\label{sec:refs}

\end{document}